\documentclass[a4paper]{article}

\usepackage{INTERSPEECH2019}

\title{Non-Parallel Voice Conversion with Cyclic Variational Autoencoder}
\name{Patrick Lumban Tobing$^1$, Yi-Chiao Wu$^1$, Tomoki Hayashi$^1$, Kazuhiro Kobayashi$^2$, Tomoki Toda $^2$}
\address{
  $^1$Graduate School of Information Science, Nagoya University, Japan\\
  $^2$Information Technology Center, Nagoya University, Japan}
\email{\{patrick.lumbantobing, yichiao.wu, hayashi.tomoki\}@g.sp.m.is.nagoya-u.ac.jp,\\
kobayashi.kazuhiro@g.sp.m.is.nagoya-u.ac.jp, tomoki@icts.nagoya-u.ac.jp}

\begin{document}

\maketitle
\begin{abstract}
In this paper, we present a novel technique for a non-parallel voice conversion (VC) with the use of
cyclic variational autoencoder (CycleVAE)-based spectral modeling. In a variational autoencoder
(VAE) framework, a latent space, usually with a Gaussian prior, is used to encode a set of input
features. In a VAE-based VC, the encoded latent features are fed into a decoder, along with
speaker-coding features, to generate estimated spectra with either the original
speaker identity (reconstructed) or another speaker identity (converted). Due to the non-parallel
modeling condition, the converted spectra can not be directly optimized, which heavily degrades
the performance of a VAE-based VC. In this work, to overcome this problem, we propose to use CycleVAE-based
spectral model that indirectly optimizes the conversion flow by recycling the converted features back into the
system to obtain corresponding cyclic reconstructed spectra that can be directly optimized. The cyclic
flow can be continued by using the cyclic reconstructed features as input for the next cycle. The experimental
results demonstrate the effectiveness of the proposed CycleVAE-based VC, which yields higher accuracy of converted
spectra, generates latent features with higher correlation degree, and significantly improves the quality
and conversion accuracy of the converted speech.
\end{abstract}

\section{Introduction}
Using a voice conversion (VC) system, voice characteristics of a source speaker can be transformed into that of a
desired target speaker, while keeping the linguistic contents intact. Such transformation can be achieved, for
example, by performing statistical conversion of spectral envelope parameters of the vocal tract spectrum, and a
proper alteration of excitation features, such as fundamental frequency ($F_0$). Within two decades, many speech
applications have been realized by employing the VC framework, such as creation of speech database with various
voice characteristics \cite{Kain98}, singing voice conversion \cite{Kobayashi18}, recovery of
impaired speech signal \cite{Kain07,Tanaka13}, expressive speech synthesis \cite{Inanoglu09,Turk10},
body-conducted speech processing \cite{Subramanya08,Toda12}, and articulatory controllable speech modification
\cite{Tobing17}. For flexible development of related applications, it is important to consider a VC technique
that can be realized using easily available speech data.

There are two main VC frameworks, non-parallel VC and parallel VC. In the non-parallel VC, it is not
straightforward to measure the correspondence between source spectral features and the target spectral features,
due to the non-existence of paired utterances. On the other hand, in a parallel VC \cite{Stylianou98,Toda07},
because of the availability of the paired utterances, their correspondence can be directly achieved by performing
time-alignment, such as with dynamic-time-warping (DTW) algorithm. However, not all of the time a proper parallel
dataset, i.e., where the source and the target speakers utter the same set of sentences, can be collected for the
development of a VC system. Consequently, as our main focus in this work, a consideration for a reliable
non-parallel VC using data-driven statistical modeling would be highly beneficial for
real-life applications.

Indeed, the challenge in developing the non-parallel spectral conversion model has attracted many works within
the recent years, such as: with the use of clustered spectral matching algorithms \cite{Erro10b,Benisty14};
with adaptation/alignment of speaker model parameters \cite{Zhang08,Song14}; with restricted Boltzmann machine
\cite{Nakashika16}; with generative adversarial networks (GAN)-based methods \cite{Fang18,Kameoka18}; and with
variational autoencoder (VAE)-based frameworks \cite{Hsu16,Hsu17,Saito18,Kameoka18b}. In this work, we focus on
the use of VAE-based system, due to its potential in employing latent space to represent common hidden aspects
of speech signal, between different speakers, e.g., phonetical attributes. Further, its implementation can be
flexibly realized through any network architectures, such as with convolutional or recurrent models.

In a VAE framework \cite{Kingma13}, a latent space, usually with a Gaussian prior, is used for encoding a set
of input features. In a VAE-based VC \cite{Hsu16}, additional speaker-coding features are used, alongside the
encoded latent features, to reconstruct the spectral features in the generation phase. Speaker-code associated
with the source (original) speaker is used to estimate the reconstructed spectra, while speaker-code associated
with a desired target speaker is used to estimate converted spectra. However, due to the non-parallel condition,
the spectral model parameters are optimized with respect only to the reconstructed spectra. Hence, because of
the only reliance in speaker-code capability to disentangle speaker identity, the performance of a conventional
VAE-based VC is still insufficient.

In this paper, to improve VAE-based VC, we propose to use cycle-consistent mapping flow \cite{Wang19}, i.e.,
CycleVAE-based VC, that indirectly optimizes the conversion flow by recycling the converted spectral features.
Specifically, in the proposed CycleVAE, the converted features are fed-back into the system to generate
corresponding cyclic reconstructed spectra that can be directly optimized. The cyclic flow can, then, be continued
by feeding the cyclic reconstructed features back into the system. Therefore, the conversion flow, i.e., the
estimation of converted spectra, is indirectly considered in the computation of both the reconstruction losses
and the regularizations of latent space. In the experiments, it has been demonstrated that the proposed
CycleVAE-based VC shows higher correlation degree of latent features, i.e., more similar latent attributes
between different speakers (possibly within phonetical space), and higher accuracy of converted spectra.
Perceptual evaluation also shows significant improvements in both quality and accuracy of converted speech,
especially when the speaker identities are considerably distant, such as in cross-gender conversions.

\section{Conventional VAE-based VC}
The flow of conventional VAE-based VC is illustrated by the upper part of Fig.~\ref{fig:flow_cycvae}. Let
$\vec{X}_t=[\vec{e}_t^{(x)^{\top}},\vec{s}_t^{(x)^{\top}}]^{\top}$,
$\vec{e}^{(x)}_t=[e_t^{(x)}(1),\dotsc,e_t^{(x)}(D_e)]^{\top}$, and
$\vec{s}^{(x)}_t=[s_t^{(x)}(1),\dotsc,s_t^{(x)}(D_s)]^{\top}$ be the $D_e+D_s$, $D_e$, and $D_s$-dimensional
feature vectors of the input, the excitation, and the spectra, respectively, at frame $t$. In the training phase,
given a set of network parameters $\{\vec{\theta},\vec{\phi}\}$, a sequence of input features
$\vec{X}=[\vec{X}_1^{\top},\dotsc,\vec{X}_T^{\top}]^{\top}$ and time-invariant $D_c$-dimensional source
speaker-code features $\vec{c}^{(x)}$ \cite{Hsu16}, a set of updated network parameters
$\{\hat{\vec{\theta}},\hat{\vec{\phi}}\}$ is estimated by maximizing the variational lower bound function
\cite{Kingma13} as follows:
\begin{equation}
  \{\hat{\vec{\theta}},\hat{\vec{\phi}}\} = \argmax_{\vec{\theta},\vec{\phi}} \sum_{t=1}^{T}\mathcal{L}(\vec{\theta},\vec{\phi},\vec{X}_t,\vec{c}^{(x)}),
\label{eq:upd_param}
\end{equation}
where
\begin{align}
  \!\mathcal{L}(\vec{\theta}\!,\vec{\phi},\!\vec{X}_t,\vec{c}^{(x)})\!&=
        \!-\!D_{KL}(q_{\vec{\phi}}(\vec{z}_t|\vec{X}_t)||p_{\vec{\theta}}(\vec{z}_t)) \notag \\
        &\phantom{,,,}+\!\mathbb{E}_{q_{\vec{\phi}(\vec{z}_t|\vec{X}_t)}}
            [\log{p_{\vec{\theta}}(\vec{s}^{(x)}_t|\vec{z}_t,\!\vec{c}^{(x)})}],\\
  q_{\vec{\phi}}(\vec{z}_t|\vec{X}_t)\!&=\!\mathcal{N}(\vec{z}_t;\!f^{(\mu)}_{\vec{\phi}}(\vec{X}_t),\diag(f^{(\sigma)}_{\vec{\phi}}(\vec{X}_t)^2)),\!\!\!\\
  p_{\vec{\theta}}(\vec{s}^{(x)}_t|\vec{z}_t,\!\vec{c}^{(x)})\!&\approx\!
    \mathcal{N}(\vec{s}^{(x)}_t;g_{\vec{\theta}}(\hat{\vec{z}}^{(x)}_t,\vec{c}^{(x)}),\vec{I}),\\
  \!\!\!\!\hat{\vec{z}}^{(x)}_t\!=\!
        f^{(\mu)}_{\vec{\phi}}(\vec{X}_t)\!&+\!f^{(\sigma)}_{\vec{\phi}}(\vec{X}_t)\odot\vec{\epsilon}
    \phantom{,,}\text{s. t. }\phantom{,}\vec{\epsilon}\sim\mathcal{N}(\vec{0},\vec{I}).
\label{eq:vlb}
\end{align}
$\vec{z}_t$ denotes a $D_z$-dimensional latent feature vector, $f_{\vec{\phi}}(\cdot)$ denotes an encoder
network, $g_{\vec{\theta}}(\cdot)$ denotes a decoder network, $\odot$ denotes an element-wise product, and
$\mathcal{N}(;\vec{\mu},\vec{\Sigma})$ is for a Gaussian distribution with mean vector $\vec{\mu}$ and
covariance matrix $\vec{\Sigma}$.


Therefore, the reconstructed source spectra feature vector
$\hat{\vec{s}}_t^{(x)}$, i.e., estimated spectra with the same speaker characteristics as the input source
speaker, is given by
\begin{equation}
  \hat{\vec{s}}_t^{(x)} = g_{\vec{\theta}}(\hat{\vec{z}}^{(x)}_t,\vec{c}^{(x)}).
\label{eq:gen_rec}
\end{equation}
On the other hand, the converted source-to-target spectra $\hat{\vec{s}}_t^{(y|x)}$,
i.e., estimated spectra with the voice characteristics of a desired target speaker, is given by
\begin{equation}
  \hat{\vec{s}}_t^{(y|x)} = g_{\vec{\theta}}(\hat{\vec{z}}^{(x)}_t,\vec{c}^{(y)}),
\label{eq:gen_cv}
\end{equation}
where $\vec{c}^{(y)}$ denotes the time-invariant $D_c$-dimensional target speaker-code features \cite{Hsu16}.
In this paper, we use not only source, but also target speakers as input in training. In order to use
the corresponding target speaker as the input speaker, i.e., optimization of reconstructed target spectra and/or
performing target-to-source conversion, the notations of $x$ and $y$, in Eqs.~(1)--(7), are swapped with each
other. Though, the performance of VAE-based VC is noticeably insufficient because the conversion
flow is not considered in the parameter optimization.

\section{Proposed CycleVAE-based VC}
\begin{figure}[!t]
  \centering
  \includegraphics[width=\linewidth]{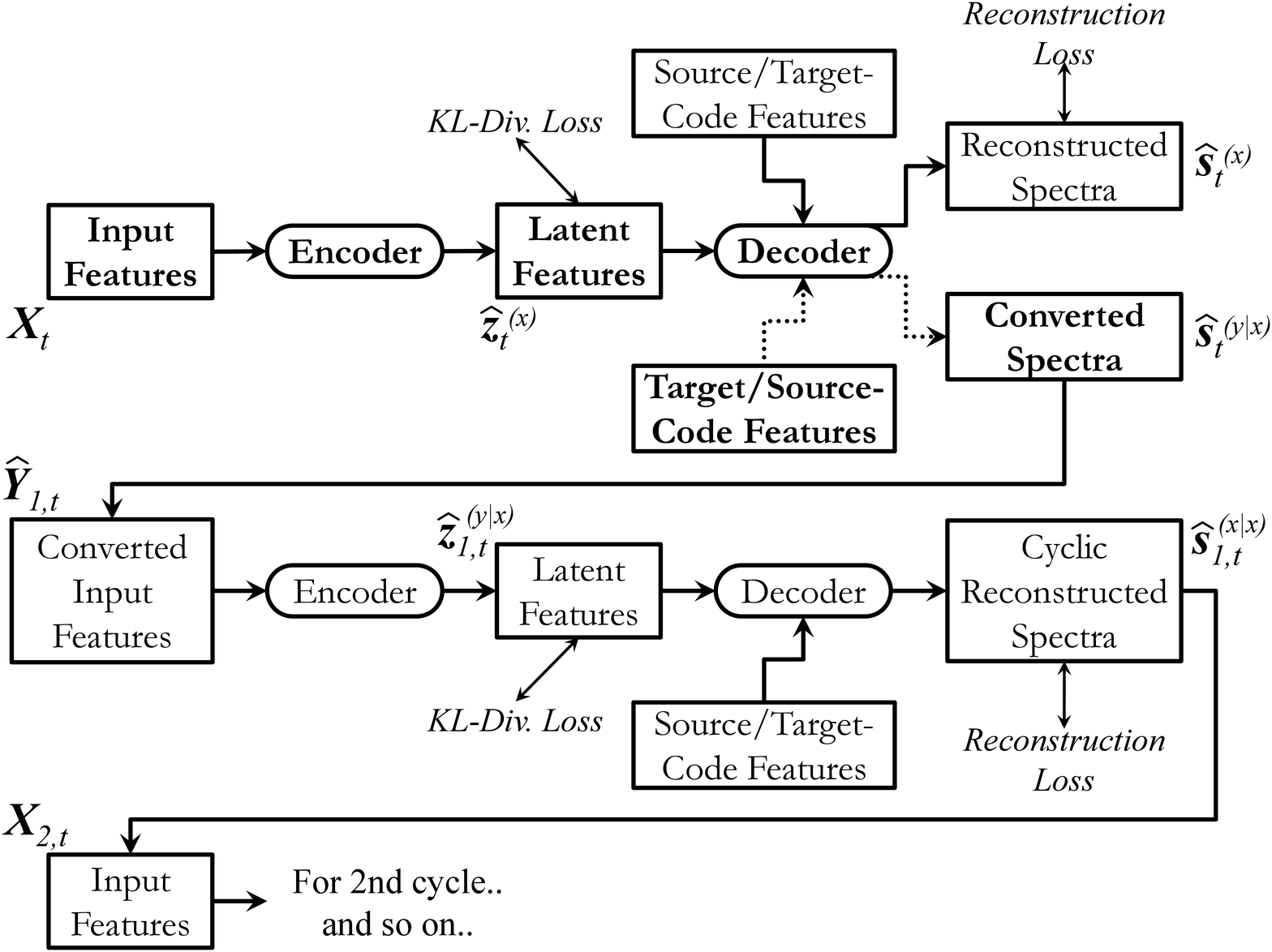}
  \caption{Flow of the conventional VAE-based (upper-part) and the proposed CycleVAE-based (whole diagram) VC.
  Converted input features include converted excitation features, such as linearly transformed $F_0$ values. One
  full-cycle includes the estimation of both reconstructed and cyclic reconstructed spectra. Each of encoder
  and decoder networks are shared for all cycles.}
  \label{fig:flow_cycvae}
  \vspace{-2.35mm}
\end{figure}

In this paper, to improve the VAE-based VC, as illustrated in Fig.~\ref{fig:flow_cycvae}, we propose
CycleVAE, which is capable of recycling the converted spectra back into the system, so that the
conversion flow is indirectly considered in the parameter optimization. A similar idea has also been proposed
as a cycle-consistent flow in a self-supervised method for visual correspondence \cite{Wang19}.

In the proposed CycleVAE-based VC, the parameter optimization is defined as follows:
\begin{equation}
    \{\hat{\vec{\theta}},\hat{\vec{\phi}}\} = \argmax_{\vec{\theta},\vec{\phi}} \sum_{t=1}^{T}\mathcal{L}(\vec{\theta},\vec{\phi},\vec{X}_t,\vec{c}^{(x)},\vec{c}^{(y)}),
\label{eq:upd_param_cyc}
\end{equation}
where
\begin{align}
  \!&\mathcal{L}(\vec{\theta}\!,\vec{\phi},\!\vec{X}_t,\vec{c}^{(x)},\vec{c}^{(y)})\!=\sum_{n=1}^{N}
        \!-D_{KL}(q_{\vec{\phi}}(\vec{z}_{n,t}|\vec{X}_{n,t})||p_{\vec{\theta}}(\vec{z}_t)) \notag \\
   &\phantom{,,,,,,,,}-\!D_{KL}(q_{\vec{\phi}}(\vec{z}_{n,t}|\hat{\vec{Y}}_{n,t})||p_{\vec{\theta}}(\vec{z}_t)) \notag \\
   &\phantom{,,,,,,,,}+\!\mathbb{E}_{q_{\vec{\phi}(\vec{z}_t|\vec{X}_t)}}
            [\log{p_{\vec{\theta}}(\vec{s}^{(x)}_{n,t}=\vec{s}^{(x)}_t|\vec{z}_{n,t},\!\vec{c}^{(x)})}] \notag \\
   &\phantom{,,,,,,,,}+\mathbb{E}_{q_{\vec{\phi}(\vec{z}_t|\hat{\vec{Y}}_t)}}
            [\log{p_{\vec{\theta}}(\vec{s}^{(x|x)}_{n,t}=\vec{s}^{(x)}_t|\vec{z}_{n,t},\!\vec{c}^{(x)})}],\\
  &q_{\vec{\phi}}(\vec{z}_{n,t}|\hat{\vec{Y}}_{n,t})\!=\!\mathcal{N}(\vec{z}_{n,t};
    f^{(\mu)}_{\vec{\phi}}(\hat{\vec{Y}}_{n,t})\!\!,\diag(f^{(\sigma)}_{\vec{\phi}}(\hat{\vec{Y}}_{n,t})^2)),\!\!\!\\
  &p_{\vec{\theta}}(\vec{s}^{(x|x)}_{n,t}|\vec{z}_{n,t},\!\vec{c}^{(x)})\approx
    \mathcal{N}(\vec{s}^{(x)}_t;g_{\vec{\theta}}(\hat{\vec{z}}^{(y|x)}_{n,t},\vec{c}^{(x)}),\vec{I}),\\
  &\hat{\vec{z}}^{(y|x)}_{n,t}\!=\!
        f^{(\mu)}_{\vec{\phi}}(\hat{\vec{Y}}_{n,t})\!+\!f^{(\sigma)}_{\vec{\phi}}(\hat{\vec{Y}}_{n,t})\!\odot\vec{\epsilon}
    \phantom{,}\text{s. t. }\vec{\epsilon}\sim\mathcal{N}(\vec{0},\vec{I}),\!\!\!
\label{eq:vlb_cyc}
\end{align}
where $\vec{s}^{(x)}_{n,t}$ and $\vec{s}^{(x|x)}_{n,t}$ are random variables, $\vec{s}^{(x)}_t$ is
an observed value, and
\begin{align}
    \hat{\vec{Y}}_{n,t} &= [\hat{\vec{e}}^{(y|x)^{\top}}_t,\hat{\vec{s}}^{(y|x)^{\top}}_{n,t}]^{\top},\\
    \hat{\vec{s}}^{(y|x)}_{n,t} &= g_{\vec{\theta}}(\hat{\vec{z}}^{(x)}_{n,t},\vec{c}^{(y)}),\\
    \hat{\vec{s}}^{(x)}_{n,t} &= g_{\vec{\theta}}(\hat{\vec{z}}^{(x)}_{n,t},\vec{c}^{(x)}),\\
    \vec{X}_{n,t} &= [\vec{e}^{(x)^{\top}}_t,\hat{\vec{s}}^{(x|x)^{\top}}_{n-1,t}]^{\top},\\
    \hat{\vec{s}}^{(x|x)}_{n,t} &= g_{\vec{\theta}}(\hat{\vec{z}}^{(y|x)}_{n,t},\vec{c}^{(x)}).
\label{eq:cvfeat_cyc}
\end{align}
The index of the $n$-th cycle is denoted as $n$. The total number of cycle is $N$.
$\hat{\vec{Y}}_{n,t}$ denotes the converted input features at $n$-th cycle,
$\hat{\vec{e}}^{(y|x)}_t$ denotes the converted source-to-target excitation features, e.g., linearly transformed
$F_0$, $\hat{\vec{s}}^{(x|x)}_{n,t}$ denotes the cyclic reconstructed spectra at $n$-th cycle, and
at $n=1$, $\hat{\vec{s}}^{(y|x)}_{1,t}=\hat{\vec{s}}^{(y|x)}_t$,
$\hat{\vec{s}}^{(x)}_{1,t}=\hat{\vec{s}}^{(x)}_t$, $\hat{\vec{z}}^{(x)}_{1,t}=\hat{\vec{z}}^{(x)}_t$
and $\vec{X}_{1,t}=\vec{X}_t$. Hence, in the proposed CycleVAE-based VC, the conversion flow is indirectly
optimized through the consideration of the converted spectra $\hat{\vec{s}}^{(y|x)}_{n,t}$ in each $n$-th cycle.

\section{Experimental Evaluation}

\begin{figure}[!t]
  \centering
  \includegraphics[width=\linewidth]{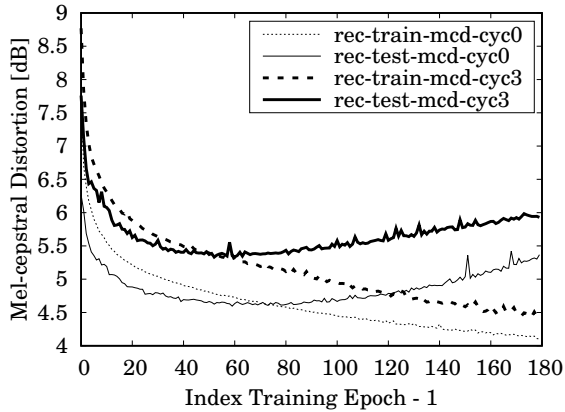}
  \caption{Mel-cepstral distortion (mcd) of reconstructed (rec) spectra, estimated using the conventional
  VAE-based (cyc0) and the proposed CycleVAE-based (cyc3) VC, during 180 training epochs, for training
  (train) and testing (test) sets. mcds were computed with only the speech frames of the input speech.}
  \label{fig:rec-mcd}
  \vspace*{-2.35mm}
\end{figure}

\subsection{Experimental conditions}

We used a subset of the Voice Conversion Challenge (VCC) 2018 \cite{Jaime18} dataset, which included four
speakers, i.e., SF1, SM1, TF1, and TM1. The speaker notations are as follows: S denotes source speaker, T denotes
target speaker, F denotes female speaker, and M denotes male speaker. The total number of utterances in the
training and the testing sets were 81 and 35, respectively. The average length per one audio sample is
about 3.5 seconds. To develop a non-parallel training dataset, the first 40 utterances were used for
corresponding source speaker, while the last 41 were for the target speaker.

WORLD \cite{Morise16} package was used to perform speech analysis. As the spectral envelope parameters, we used
the zeroth through 34$^{\text{th}}$ mel-cepstrum coeficients converted from the spectral envelope, which was
extracted frame-by-frame. As the excitation features, we used log-scaled of continuous $F_0$ also including an
unvoiced/voiced binary decision feature, and 2-dimensional aperiodicity coding coefficients. To perform
excitation conversion, mean and variance transformation \cite{Toda07} was performed with respect to the
log-scaled $F_0$ values. The sampling rate of the speech signal was 22,050~kHz. The number of FFT points was
1024. The frame shift was set to 5~ms.

To develop the spectral networks, we used a recurrent neural network (RNN)-based model, which was as follows:
dilated convolutional layers were used, to capture the context of -4/+4 input frames, with a kernel size of 3 and
2 layers of 1 and 3 dilation, respectively; gated recurrent unit (GRU) \cite{Cho14} was used with 1024 hidden
units and 1 hidden layer; a linear output layer was used; output frame was also fed-back into GRU. Fixed
normalization and denormalization layers were used before convolutional and after output layers, respectively,
that were set with the statistics of training data. Dropout \cite{Srivastava14} layers were used with 0.5
probability after convolutional and GRU layers. Network parameters are initialized with Glorot \cite{Glorot10}
method, and optimized using Adam \cite{Kingma14} with 0.0001 learning rate. A batch-frame size of 80 was used.

Four one-to-one spectral models were developed for each of the conventional VAE- and the proposed CycleVAE-based
VC, with respect to the four corresponding speaker pairs, i.e., SF1-TF1, SF1-TM1, SM1-TF1, and SM1-TM1. To code
the speaker identity, a binary decision value was used. Search of hyperparameters was conducted by varying the
number of latent dimensions to 8, 16, 32, 50, and 64, and the number of cycles $N$, in
Eq.~\eqref{eq:vlb_cyc}, to 1, 2, 3, 4, and 5. The optimum number of latent dimensions for both VAE and CycleVAE
was 16. The optimum number of cycles for CycleVAE was 3. Objective evaluation was performed to measure the
accuracy of the reconstructed and the converted spectra, and the degree of latent features correlation. Another
RNN-based parallel spectral conversion models were developed as the upper bound in measuring conversion accuracy.
Subjective evaluation was performed to perceptually measure the quality and the accuracy of converted speech
between conventional VAE and proposed CycleVAE
\footnote{Implementation is being made available at \url{https://github.com/patrickltobing/cyclevae-vc}}.

\begin{figure}[!t]
  \centering
  \includegraphics[width=\linewidth]{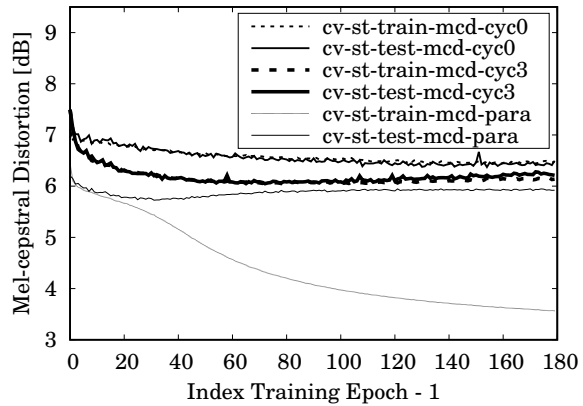}
  \caption{Mel-cepstral distortion (mcd) of converted source-to-target (cv-st) spectra, estimated using the
  conventional VAE-based (cyc0) and the proposed CycleVAE-based (cyc3) VC, during 180 training epochs, for
  training (train) and testing (test) sets. mcds were computed, through DTW alignment, with only the speech
  frames of corresponding source and target speech.}
  \label{fig:cv-st-mcd}
\end{figure}

\begin{table}[t]
\footnotesize
\centering
\caption{Mel-cepstral distortion (MCD) of converted spectra (Cv) and GV-postfiltered \cite{Toda07} converted
spectra (PF) with the conventional VAE, the proposed CycleVAE, and parallel spectral modeling as the
lower bound, for each speaker-pair conversions. (S: source speaker; T: target speaker; F: female speaker;
M: male speaker; Init. denotes the initial MCD values.)}
\begin{tabular}{@{}c|c|cc|cc|cc@{}}
  \toprule
  \multirow{2}{*}{\textbf{MCD [dB]}} & \multirow{2}{*}{\textbf{Init.}} & \multicolumn{2}{c|}{\textbf{VAE}} & \multicolumn{2}{c|}{\textbf{CycleVAE}} & \multicolumn{2}{c}{\textbf{Parallel}} \\
   & & \textbf{Cv} & \textbf{PF} & \textbf{Cv} & \textbf{PF} & \textbf{Cv} & \textbf{PF} \\ \midrule
  \textbf{SF1-TF1} & 8.18 & 6.41 & 6.95 & \textbf{6.24} & \textbf{6.78} & 5.92 & 6.42 \\
  \textbf{SF1-TM1} & 8.73 & 6.49 & 7.03 & \textbf{5.97} & \textbf{6.49} & 5.60 & 6.03 \\
  \textbf{SM1-TF1} & 9.06 & 6.83 & 7.42 & \textbf{6.29} & \textbf{6.78} & 6.00 & 6.43 \\
  \textbf{SM1-TM1} & 7.68 & 5.74 & 6.15 & \textbf{5.71} & \textbf{6.10} & 5.36 & 5.72 \\ \bottomrule
\end{tabular}
\label{tab:cv-pf-mcd}
\vspace*{-2.35mm}
\end{table}

\subsection{Objective evaluation}

Mel-cepstral distortion (MCD) \cite{Toda07} was used to measure the accuracy of both the reconstructed and
the converted spectra. Their values are respectively charted, during 180 training epochs, in
Figs.~\ref{fig:rec-mcd} and \ref{fig:cv-st-mcd}. It can be observed that the proposed CycleVAE-based VC yields
higher accuracy of converted spectra and lower accuracy of reconstructed spectra compared to the conventional
VAE. This trend is somewhat inline with \cite{Chorowski19}, where reconstruction performance is not a proper
measure for a better disentanglement of speaker identity (or for better conversion performance). Moreover, MCD
values of converted spectra were also computed after applying global variance (GV)-postfilter \cite{Toda07}, as
given in Table ~\ref{tab:cv-pf-mcd}. The result shows that the proposed CycleVAE is more suited to additional
postfiltering phase compared to the conventional VAE, especially when the speaker identities are
considerably distant.

\begin{figure}[!t]
  \centering
  \includegraphics[width=\linewidth]{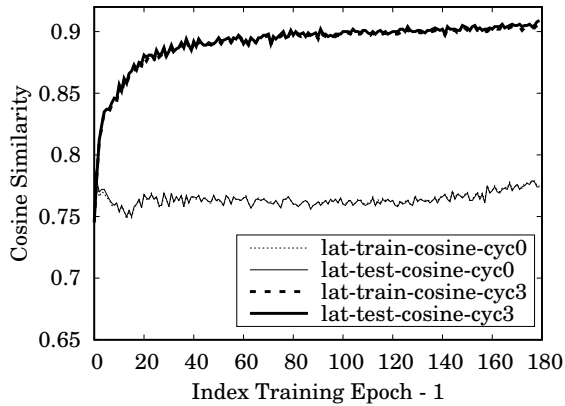}
  \caption{Cosine similarity (cosine) between latent features of corresponding source and target speech, encoded
  with the conventional VAE-based (cyc0) and the proposed CycleVAE-based (cyc3) VC, during 180 training epochs,
  for training (train) and testing (test) sets. cosines were computed, through DTW alignment, with only the
  speech frames of source and target speech.}
  \label{fig:lat-cosine}
  \vspace*{-2.35mm}
\end{figure}

To measure the condition of the latent features, we computed the cosine similarities between the latent
features of the source and of the target speaker within the same utterances, which were charted during 180
training epochs, as in Fig.~\ref{fig:lat-cosine}. It can be clearly seen that the proposed CycleVAE-based VC
generates latent features with higher correlation degree compared to conventional VAE. As studied in
\cite{WNHsu17b}, higher cosine similarities would be produced by latent attributes that represent either equal
phonetic space or equal speaker identities. Hence, CycleVAE is more likely to give latent representations that
are closer to phonetic domain due to different speaker identities.


\subsection{Subjective evaluation}

Perceptual evaluation was performed to compare the quality and the conversion accuracy of converted speech,
between the conventional VAE- and the proposed CycleVAE-based VC, by conducting two forced-choice preference
tests. In the quality preference test, each listener was presented with two audio stimuli at a time,
and was asked to choose a prefered audio by considering both speech naturalness and intelligibility. In the
similarity preference test, i.e., to measure the conversion accuracy, each listener was given two audio stimulis,
and a reference audio with different utterance, then, was asked to choose a prefered audio that has the closer
speaker characteristics to the reference speaker. The numbers of distinct utterances in quality and similarity
tests were 6 and 5, respectively, which were randomly chosen from the testing set. Converted speech using
parallel spectral models were also included. GV-postfiltered converted spectra was used. The number of listeners
was 10.

The results of quality and similarity preference tests are given in Tables ~\ref{tab:qpref_res} and
\ref{tab:spref_res}, respectively. These results show that the proposed CycleVAE-based VC significantly
improves the overall quality and accuracy of converted speech, especially for cross-gender (SF1-TM1, SM1-TF1)
conversions, compared to conventional VAE. Their performances for same-gender conversions are statistically
similar. This tendency is inline with the objective measurements shown in Table~\ref{tab:cv-pf-mcd}, where the
conventional VAE-based VC suffers from degradation in cross-gender conversions and the CycleVAE significantly
improves them. All audio samples and complete perceptual results can be accessed at \url{http://bit.ly/2Wg3oIt}.

\begin{table}[!t]
\centering
\caption{Result of preference test on speech quality for all, same-gender (S-Gender), and cross-gender
(X-Gender) conversion categories using the conventional VAE and the proposed CycleVAE-based VC. CI denotes
the 95$\%$ confidence interval of the sample mean. p-values were computed using the two-tailed Mann--Whitney
U-test with $\alpha <$~0.05. Bold indicates statistically significant better scores.}
\begin{tabular}{@{}c|c|c|c|c@{}}
  \toprule
  \textbf{\begin{tabular}[c]{@{}c@{}}\textbf{Quality}\\\textbf{Preference}\end{tabular}} & \textbf{VAE} & \textbf{CycleVAE} & \textbf{CI} & \textbf{p-value} \\ \midrule
  \textbf{All} & 40.83\% & \textbf{59.17\%} &$\pm$6.27\% & 6.01e-05 \\
  \textbf{S-Gender} & 52.50\% & 47.50\% & $\pm$9.07\% & 4.40e-01 \\
  \textbf{X-Gender} & 29.17\% & \textbf{70.83\%} & $\pm$8.25\% & 1.18e-10 \\ \bottomrule
\end{tabular}
\label{tab:qpref_res}
\end{table}

\begin{table}[t]
\centering
\caption{Result of preference test on speaker similarity (Spk. Sim.) for all, same-gender (S-Gender), and
cross-gender (X-Gender) conversion categories using the conventional VAE and the proposed CycleVAE-based VC. CI
denotes the 95$\%$ confidence interval of the sample mean. p-values were computed using the two-tailed
Mann--Whitney U-test with $\alpha <$~0.05. Bold indicates statistically significant better scores.}
\begin{tabular}{@{}c|c|c|c|c@{}}
  \toprule
  \textbf{\begin{tabular}[c]{@{}c@{}}\textbf{Spk. Sim.}\\\textbf{Preference}\end{tabular}} & \textbf{VAE} & \textbf{CycleVAE} & \textbf{CI} & \textbf{p-value} \\ \midrule
  \textbf{All} & 39.00\% & \textbf{61.00\%} &$\pm$6.82\% & 1.11e-05 \\
  \textbf{S-Gender} & 46.00\% & 54.00\% & $\pm$9.94\% & 2.59e-01 \\
  \textbf{X-Gender} & 32.00\% & \textbf{68.00\%} & $\pm$9.30\% & 3.81e-07 \\ \bottomrule
\end{tabular}
\label{tab:spref_res}
\vspace*{-2.35mm}
\end{table}

\section{Conclusions}

We have presented a novel framework to improve conventional VAE, for a non-parallel VC, by using a
cycle-consistent flow, i.e., the proposed CycleVAE. Specifically, the converted spectra, which is not
directly optimized, is recycled back into the system, to generate cyclic reconstructed spectra that can be
directly optimized. The cyclic flow can be continued by feeding the cyclic reconstructed features back into
the system. The experimental results demonstrate that the proposed CycleVAE-based VC yields higher correlation
degree of latent features and more accurate converted spectra, while significantly improves the quality
and conversion accuracy of the converted speech. Future work includes development of many-to-many VC, and
incorporates the use of discrete latent space \cite{Oord17}, better prior \cite{WNHsu17}, i-vector
\cite{Kinnunen17b}, additional classifier network \cite{Kameoka18b}, and neural waveform generator
\cite{Oord16} to produce naturaly sounding converted speech \cite{Tobing19} with the proposed
CycleVAE.

\section{Acknowledgements}

This work was partly supported by JST, PRESTO Grant Number JPMJPR1657, and JSPS KAKENHI Grant Number JP17H06101.

\bibliographystyle{IEEEtran}

\bibliography{cycvae}

\end{document}